# Software Complex for the Numerical Solution of the Isotropic Imaginary-Axis Eliashberg Equations

**RENAT SH. IKHSANOV,[1,*] EVGENY A. MAZUR,[2,3] MAXIM YU. KAGAN,[1,4*]**

[1]*National Research University Higher School of Economics, Miasnitskaya Ulitsa 20, 101000 Moscow, Russia*
[2]*National Research Nuclear University MEPhI, 115409 Moscow, Russia*
[3]*Kurchatov Institute National Research Center, 123122, Moscow, ul. Academician Kurchatov, 1, Russia*
[4]*P.L. Kapitza Institute for Physical Problems, Russian Academy of Sciences, Kosygin st.2, 119334, Moscow, Russia*

*Corresponding author: rihsanov@hse.ru, kagan@kapitza.ras.ru*

**At present, the task of searching for compounds with a high superconducting transition temperature $T_C$ is a very relevant scientific direction. Usually, the calculation of $T_C$ is carried out by numerically solving the system of Eliashberg equations. In this paper, a set of programs for solving this system written in various forms on the imaginary axis is presented. As an example of the developed methods applications, calculations results of $T_C$ and thermodynamic properties of the metallic hydrogen I41/AMD phase and some other substances under high pressure are presented.**

## 1. Introduction.

The system of Eliashberg equations is a system of nonlinear integral equations that describes the phase transition of metal from a normal state to a superconducting one [1-3]. The ability to predict the superconducting transition temperature of a metal and its thermodynamic properties is very important for many practical applications. However, this system does not have an analytical solution in the general case, and, accordingly, numerical methods are needed to solve it. Such methods are developed by various authors (see, for example, [4-8]) for various forms of the Eliashberg system of equations. However, in these works, the emphasis is on the physical results obtained by solving the Eliashberg equations, and not on numerical procedures. In our work, we focus on numerical procedures, and the results obtained for various substances subjected to high pressures are used as an illustration of the developed numerical methods.

In our work, we use the system of Eliashberg equations for an isotropic superconductor, written on the imaginary axis. This form is the most convenient for numerical analysis because, firstly, only finite sums over Matsubara frequencies appear in this system (not integrals like in the real-axis version of this system), and secondly, all dependent variables of the system are real-valued functions.

The software package is implemented on Mathcad and Matlab platforms. The program interface is implemented on Mathcad platform, and the calculations are performed by a program written in Matlab language (data are transferred from the interface to the calculation module by being written to the appropriate files).

In Section 1, we present a program that implements the Allen-Dynes method [9, 10], which makes it possible to find the first approximation to $T_C$ without solving a system of nonlinear equations. Section 2 describes the program for solving the system of Eliashberg equations written for the order parameter and the electron mass renormalization function. In Section 3, we introduce an additional variable (and an additional equation) into the system – a correction to the chemical potential of electrons. Section 4 gives an example of the described methods applications to the calculations of the superconducting transition temperature and thermodynamic properties of the I41/AMD phase of metallic hydrogen and some other substances under high pressure. Earlier, we briefly announced these results in the paper [11].

## 2. The Allen-Dynes algorithm.

To calculate $T_c$ according to the Allen-Dynes algorithm [9, 10], it is required to specify the Eliashberg function $\alpha^2 F(\omega)$ (it is specified in the corresponding file in the form of a table) and the Coulomb pseudopotential of electrons $\mu^*$. The program provides the ability to set the dimension of the function $\alpha^2 F(\omega)$ argument in the three most common options: THz, eV, Ry.

For the calculation, a square matrix $K$ of the size $N \times N$ is formed:

$$K_{mn} = \lambda(|m-n|) + \lambda(m+n-1) - 2\mu^* - \delta_{mn} \times \left[ 2m - 1 + \lambda(0) + 2\sum_{l=1}^{m-1} \lambda(l) \right], \quad (1)$$

where

$$\lambda(l) = 2\int_0^{\omega_{max}} \frac{\alpha^2 F(\omega)\cdot\omega}{\omega^2 + (2\pi kTl)^2}d\omega, \qquad (2)$$

$\delta_{mn}$ is the Kronecker symbol, $m,n = \overline{1,N}$, $\omega_{max}$ is the maximum frequency in the function $\alpha^2 F(\omega)$ spectrum, $k$ is the Boltzmann's constant, $T$ is the absolute temperature. The frequency in formula (2) and further is expressed in the energy units. Note that in the original papers [9, 10], a different index numbering is used: $m,n = \overline{0,N}$, so formula (1) in [9, 10] has a slightly different form than in our work. Our numbering is more convenient for programming languages where array numbering starts from 1 (for example, Matlab).

Integral (2) is calculated numerically. It is most optimal to calculate it on the grid on which the Eliashberg function $\alpha^2 F(\omega)$ is given.

Let $\rho(T)$ be the maximum eigenvalue of the matrix $K$ formed for the temperature $T$. Then $\rho(T) = 0$ for $T \geq T_c$ [9, 10]. This fact provides a way to determine $T_c$: the program cycles through the temperature with a selected step $\Delta T$ ($\Delta T > 0$), starting from a certain temperature $T_0$ ($T_0 < T_c$), until $\rho(T)$ is equal to zero. Accordingly, $\Delta T$ is an accuracy of determining the temperature of the superconducting transition by this method. The choice of $N$ is made for the reason of $\rho(T)$ independence from $\Delta T$.

In the work [10], the following selection criterion for $N$ is proposed (it is implemented in our program):

$$\omega_N > 8\langle\omega^2\rangle^{1/2} = 8\left(\frac{2}{\lambda(0)}\int_0^{\omega_{max}}\alpha^2 F(\omega)\cdot\omega d\omega\right)^{1/2}, \qquad (3)$$

where $\omega_N$ is the Matsubara frequency with number $N$. The Matsubara frequency with the number $j$ ($j \in \mathbb{Z}$) is determined by the formula

$$\omega_j = \pi(2j+1)kT. \qquad (4)$$

The program that implements the Allen-Dynes method is included in our software package because of its simplicity and reliability. The fact is that in this method there is no need to directly solve the system of nonlinear integral equations, therefore there is no problem of initial approximation to the solution choice, and the problem of the algorithm's convergence. The disadvantages of this method include not always acceptable accuracy ($\pm 5$ K) and the impossibility of obtaining the magnitude of the superconducting gap at 0 K. Accordingly, this method is suitable for obtaining a good approximation to $T_c$ for subsequent calculations by more resource-consuming methods.

## 3. The classical system of Eliashberg equations.

The classical system of Eliashberg equations is written with respect to a set of two dependent variables: $\{\varphi, Z\}$, where $\varphi$ is the order parameter; $Z$ is the electron mass renormalization function. These variables are functions of frequency on a discrete set of Matsubara frequencies (4). The numerical parameters of the system form the set $\{T, \mu^*, \omega_c\}$, where $T$ is the absolute temperature, $\mu^*$ is the screened Coulomb potential, and $\omega_c$ is the Coulomb interaction efficiency energy range (usually taken as $\omega_c \approx 3\omega_{max}$). The functional parameter of the system is the Eliashberg function $\alpha^2 F(\omega)$.

The system of equations has the form (see, for example, [12])

$$\begin{cases} \varphi_n = \frac{\pi}{\beta}\sum_{m=-M-1}^{M}\frac{\lambda(i\omega_n - i\omega_m) - \mu^*\theta(\omega_c - |\omega_m|)}{\sqrt{(\omega_m Z_m)^2 + \varphi_m^2}}\varphi_m \\ Z_n = 1 + \frac{\pi}{\beta}\frac{1}{\omega_n}\sum_{m=-M-1}^{M}\frac{\lambda(i\omega_n - i\omega_m)}{\sqrt{(\omega_m Z_m)^2 + \varphi_m^2}}\omega_m Z_m \end{cases}, \qquad (5)$$

where $n = \overline{-M-1, M}$, $M$ is the maximum positive Matsubara frequency number, $\theta(x) = \begin{cases} 0, x < 0 \\ 1, x \geq 0 \end{cases}$ – the Heaviside function, $\beta = 1/kT$; $\lambda(z) = 2\int_0^{\omega_{max}}\frac{\alpha^2 F(\omega)\cdot\omega}{\omega^2 - z^2}d\omega$, $z$ is purely imaginary.

From the solution of the system (5), one can determine the order parameter $\Delta_0$:

$$\Delta_0 = \varphi(\omega_0)/Z(\omega_0), \qquad (6)$$

where $\varphi(\omega_0)$ and $Z(\omega_0)$ are the components of the system (5) solution, taken at zero Matsubara frequency at the fixed temperature $T$. If the order parameter (as the solution of system (5)) is written as a function of temperature, this function will have the form $\Delta_0(T) = \begin{cases} > 0, \ T < T_c \\ 0, \ T \geq T_c \end{cases}$. This dependence gives a method for determining the superconducting transition temperature from the solution of the system (5): $T_c$ is defined as the temperature of the order parameter zeroing when the numerical procedure is moving upward in temperature (from a temperature lower than $T_c$). In addition to finding $T_c$, often of interest is $\Delta_0(0)$ – the value of the superconducting gap at 0 K. $\Delta_0(0)$ is found by extrapolating the dependence $\Delta_0(T)$ to $T = 0$ (see Fig. 1).

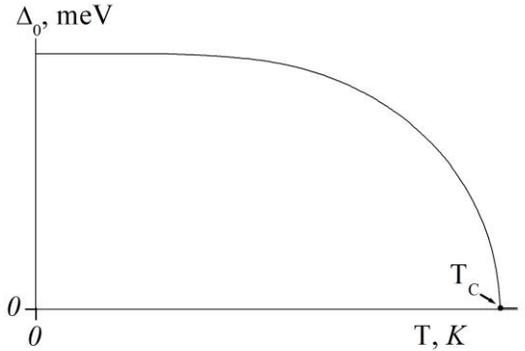

**Figure 1.** Typical dependence $\Delta_0(T)$.

In our software package, the calculation of the dependence $\Delta_0(T)$ on an arbitrary temperature interval $[T_1, T_2]$ with an arbitrary temperature step ($\Delta T$ can be either greater or less than zero) is implemented. If $\Delta T > 0$, then in the temperature cycle the number of Matsubara frequencies is kept constant (equal to the initial value). If $\Delta T < 0$, then at each step of the temperature cycle the number of Matsubara frequencies is chosen so that the Matsubara frequency with the maximum number is not less than the initially chosen one, i.e. the frequency interval does not narrow with the temperature drop. Correspondingly, in this case the number of Matsubara frequencies grows inversely proportional to the first power of the temperature and can reach hundreds or even thousands for superconductors with low $T_c$.

To shorten the calculations, we can take into account the following symmetries: $-\omega_{-n-1} = \omega_n$, $\varphi_{-n-1} = \varphi_n$, $Z_{-n-1} = Z_n$. This allows one to write sums in (5) over non-negative frequency numbers (from 0 to M):

$$\begin{cases} \varphi_n = \dfrac{\pi}{\beta} \sum_{m=0}^{M} \dfrac{\lambda_{n,m}^{(+)} - 2\mu^*\theta(\omega_c - \omega_m)}{\sqrt{(\omega_m Z_m)^2 + \varphi_m^2}} \varphi_m \\ Z_n = 1 + \dfrac{\pi}{\beta}\dfrac{1}{\omega_n} \sum_{m=0}^{M} \dfrac{\lambda_{n,m}^{(-)} \omega_m Z_m}{\sqrt{(\omega_m Z_m)^2 + \varphi_m^2}} \end{cases} \quad (6)$$

where $n = \overline{0, M}$, $\lambda_{n,m}^{(\pm)} = \lambda(n-m) \pm \lambda(n+m+1)$, $\lambda(n) = 2\int_0^{\omega_D} \dfrac{\omega \cdot \alpha^2 F(\omega)}{\omega^2 + (2\pi n k_B T)^2} d\omega$. For programming languages where array numbering starts from 1, it is more convenient to rewrite system (6) for indices $p$ and $q$, which vary from 1 to $M+1$. After the replacement $\begin{cases} n \to p : p = n+1 \\ m \to q : q = m+1 \end{cases}$ system (6) becomes:

$$\begin{cases} \varphi_p = \dfrac{\pi}{\beta} \sum_{q=1}^{M+1} \dfrac{\tilde{\lambda}_{p,q}^{(+)} - 2\mu^*\theta(\omega_c - \tilde{\omega}_q)}{\sqrt{(\tilde{\omega}_q Z_q)^2 + \varphi_q^2}} \varphi_q \\ Z_p = 1 + \dfrac{\pi}{\beta}\dfrac{1}{\tilde{\omega}_p} \sum_{q=1}^{M+1} \dfrac{\tilde{\lambda}_{p,q}^{(-)} \tilde{\omega}_q Z_q}{\sqrt{(\tilde{\omega}_q Z_q)^2 + \varphi_q^2}} \end{cases} \quad (7)$$

where $\tilde{\omega}_q = \pi k T(2q - 1)$, $\tilde{\lambda}_{p,q}^{(\pm)} = \lambda(p-q) \pm \lambda(p+q-1)$. (8)

In our software package, the classical system of Eliashberg equations is solved in the form (7).

In the program, the following empirical formulas are used as an initial approximation to the solution:

$$\begin{cases} \varphi(\omega) = \dfrac{a}{\sqrt{2\pi}\sigma} \exp\left(-\dfrac{\omega^2}{2\sigma^2}\right) \\ Z(\omega) = 1 + \dfrac{b}{\sqrt{2\pi}\sigma} \exp\left(-\dfrac{\omega^2}{2\sigma^2}\right) \end{cases} \quad (9)$$

where $a$, $b$, $\sigma$ are fitting parameters.

To apply standard numerical procedures designed to solve systems of nonlinear equations (they are available in any mathematical package) it is convenient to write system (7) in the vector form. $\mathbf{F}(\mathbf{x}) = \mathbf{0}$:

$$\begin{cases} F_1(\mathbf{x}) = 0 \\ \vdots \\ F_{2M+2}(\mathbf{x}) = 0 \end{cases} \quad (10)$$

where $\mathbf{x} = (x_i)_{i=1}^{2M+2}$, $\begin{cases} x_i = \varphi_i, \; i = \overline{1, M+1} \\ x_i = Z_{i-M-1}, \; i = \overline{M+2, 2M+2} \end{cases}$,

$F_i(\mathbf{x}) = \dfrac{\pi}{\beta} \sum_{q=1}^{M+1} \dfrac{\tilde{\lambda}_{i,q}^{(+)} - 2\mu^*\theta(\omega_c - \tilde{\omega}_q)}{\sqrt{(\tilde{\omega}_q x_{q+M+1})^2 + x_q^2}} x_q - x_i$, $i = \overline{1, M+1}$;

$F_i(\mathbf{x}) = 1 + \dfrac{\pi}{\beta}\dfrac{1}{\tilde{\omega}_{i-M-1}} \sum_{q=1}^{M+1} \dfrac{\tilde{\lambda}_{i-M-1,q}^{(-)} \tilde{\omega}_q x_{q+M+1}}{\sqrt{(\tilde{\omega}_q x_{q+M+1})^2 + x_q^2}} - x_i$, $i = \overline{M+2, 2M+2}$.

To speed up the procedure of the system (10) numerical solution and improve its convergence, one can use its Jacobian, which could be conveniently written as a block matrix.

$$J = \begin{bmatrix} A & B \\ C & D \end{bmatrix}, \quad (11)$$

where all blocks have the size $(M+1) \times (M+1)$:

$A_{ij} = \dfrac{\pi}{\beta} \dfrac{\{\tilde{\lambda}_{i,j}^{(+)} - 2\mu^*\theta(\omega_c - \tilde{\omega}_j)\}}{\left[(\tilde{\omega}_j x_{j+M+1})^2 + x_j^2\right]^{3/2}} \tilde{\omega}_j^2 x_{j+M+1}^2 - \delta_{ij}$, $i, j = \overline{1, M+1}$;

$B_{ij} = -\dfrac{\pi}{\beta} \dfrac{\{\tilde{\lambda}_{i,j-M-1}^{(+)} - 2\mu^*\theta(\omega_c - \tilde{\omega}_{j-M-1})\}}{\left[(\tilde{\omega}_{j-M-1} x_j)^2 + x_{j-M-1}^2\right]^{3/2}} \tilde{\omega}_{j-M-1}^2 x_{j-M-1} x_j$,

$i \in \overline{1, M+1}$, $j \in \overline{M+2, 2M+2}$;

$C_{ij} = -\dfrac{\pi}{\beta} \dfrac{1}{\tilde{\omega}_{i-M-1}} \dfrac{\{\tilde{\lambda}_{i,j}^{(+)} - 2\mu^*\theta(\omega_c - \tilde{\omega}_j)\}\tilde{\omega}_j x_j x_{j+M+1}}{\left[(\tilde{\omega}_j x_{j+M+1})^2 + x_j^2\right]^{3/2}}$, $i = \overline{M+2, 2M+2}$,

$j = \overline{1, M+1}$;

$D_{ij} = \dfrac{\pi}{\beta} \dfrac{1}{\tilde{\omega}_{i-M-1}} \dfrac{\{\tilde{\lambda}_{i,j-M-1}^{(+)} - 2\mu^*\theta(\omega_c - \tilde{\omega}_{j-M-1})\}\tilde{\omega}_{j-M-1} x_{j-M-1}^2}{\left[(\tilde{\omega}_{j-M-1} x_j)^2 + x_{j-M-1}^2\right]^{3/2}} - \delta_{ij}$,

$i, j = \overline{M+2, 2M+2}$.

The correctness of the above formulas for the Jacobian was checked numerically by setting the "DerivativeCheck" option in the Matlab "fsolve" function (this function solves the system (10)).

**4. The system of Eliashberg equations with a correction for the chemical potential.**

The accuracy of the superconducting transition description can be improved by introducing a correction for the electron's chemical potential into the system of Eliashberg equations. Accordingly, a new variable $\chi$ and a new equation are introduced into the system. The number of numeric parameters will also increase by 1 because the chemical potential of electrons $\mu$ is introduced. Another functional parameter $N_0(E) = N(E)/N(0)$ is also introduced, where $N(E)$ is the electron density of states. The functions from the set $\{\varphi, Z, \chi\}$ discretized on the grid of Matsubara frequencies form the variables set $\{\varphi_n, Z_n, \chi_n\}_{n=0}^{M}$ of the new system:

$$\begin{cases} \varphi_n = \dfrac{\pi}{\beta} \sum_{m=0}^{M} \{\lambda_{n,m}^{(+)} - 2\mu^*\theta(\omega_c - |\omega_m|)\} \varphi_m \cdot N_m(\varphi_m, Z_m, \chi_m) \\ Z_n = 1 + \dfrac{\pi}{\beta}\dfrac{1}{\omega_n} \sum_{m=0}^{M} \lambda_{n,m}^{(-)} \omega_m Z_m \cdot N_m(\varphi_m, Z_m, \chi_m) \\ \chi_n = -\dfrac{\pi}{\beta} \sum_{m=0}^{M} \{\lambda_{n,m}^{(+)} - 2\mu^*\theta(\omega_c - |\omega_m|)\} \cdot P_m(\varphi_m, Z_m, \chi_m) e_m(\varphi_m, Z_m) \end{cases} \quad (12)$$

where $n = \overline{0, M}$, $N_m(\varphi, Z, \chi) = \int_{-\mu}^{+\infty} \dfrac{\pi^{-1} N_0(E)}{(E+\chi)^2 + Z^2\omega_m^2 + \varphi^2} dE$,

$P_m(\varphi, Z, \chi) = \int_{-\mu}^{+\infty} \dfrac{\pi^{-1} N_0(E)(E+\chi)}{(E+\chi)^2 + Z^2\omega_m^2 + \varphi^2} dE$, $e_m(\varphi, Z) = \dfrac{\omega_m Z}{\sqrt{\omega_m^2 Z^2 + \varphi^2}}$.

If we put the electron density constant (i.e. $N_0(E) = N_0(0) = 1$), and the lower limits in the integral representations of the functions $N_m$ and $P_m$ equal to $-\infty$, then the 3rd equation of system (12) goes into $\chi_m = 0$, $\forall m$, and the other two into the corresponding equations of the previous system (7).

To write the system (12) in the vector form $F(x) = 0$, we introduce some additional notations. Let $x = (x_i)_{i=1}^{3M+3}$ be a set of the system variables, where correspondence with physical variables occurs according to formulas: $\varphi_m = x_{m+1}$, $Z_m = x_{m+M+2}$, $\chi_m = x_{m+2M+3}$, $m = \overline{0, M}$. Also we introduce functions $\tilde{N}_q$, $\tilde{P}_q$ and $\tilde{e}_q$, that differ from the corresponding functions $N_q$, $P_q$ and $e_q$ only by replacing frequencies $\omega_p$ (4) in functions without tildes with frequencies $\tilde{\omega}_p$ (8) in functions with tildes. For brevity, we denote $\mathbf{y}_q = (x_1, x_{q+M+1}, x_{q+2M+2})$ and $\mathbf{y} = (x_1, x_2, x_3)$.

In these notations, system (12) takes the form

$$F_i(\mathbf{x}) = \frac{\pi}{\beta} \sum_{q=1}^{M+1} \left\{ \tilde{\lambda}_{iq}^{(+)} - 2\mu^* \theta(\omega_c - |\tilde{\omega}_q|) \right\} x_q \cdot \tilde{N}_q(\mathbf{y}_q) - x_i, \ i = \overline{1, M+1};$$

$$F_i(\mathbf{x}) = 1 + \frac{\pi}{\beta} \frac{1}{\tilde{\omega}_{i-M-1}} \sum_{q=1}^{M+1} \tilde{\lambda}_{i-M-1,q}^{(-)} \tilde{\omega}_q x_{q+M+1} \cdot \tilde{N}_q(\mathbf{y}_q) - x_i,$$

$i = \overline{M+2, 2M+2};$

$$F_i(\mathbf{x}) = -\frac{\pi}{\beta} \sum_{q=1}^{M+1} \left\{ \tilde{\lambda}_{i-2M-2,q}^{(+)} - 2\mu^* \theta(\omega_c - |\tilde{\omega}_q|) \right\} \cdot \tilde{P}_q(\mathbf{y}_q) \tilde{e}_q(\mathbf{y}_q) - x_i,$$

$i \in \overline{2M+3, 3M+3}.$

Let us write down the Jacobian of this system left side.

$$J_{ij} = \left( \frac{\partial F_i}{\partial x_j} \right)_{i,j=1}^{3M+3}. \tag{13}$$

So, for $i \in \overline{1, M+1}$ we have:

$$J_{ij} = \frac{\pi}{\beta} \left\{ \tilde{\lambda}_{ij}^{(+)} - 2\mu^* \theta(\omega_c - |\tilde{\omega}_j|) \right\} \tilde{N}_j(\mathbf{y}_j)$$
$$+ \frac{\pi}{\beta} \sum_{q=1}^{M+1} \left\{ \left[ \tilde{\lambda}_{iq}^{(+)} - 2\mu^* \theta(\omega_c - |\tilde{\omega}_q|) \right] \frac{\partial \tilde{N}_q}{\partial x_j}(\mathbf{y}_q) \right\} - \delta_{ij}, \ j \in \overline{1, M+1};$$

$$J_{ij} = \frac{\pi}{\beta} \sum_{q=1}^{M+1} \left\{ \left[ \tilde{\lambda}_{iq}^{(+)} - 2\mu^* \theta(\omega_c - |\tilde{\omega}_q|) \right] x_q \frac{\partial \tilde{N}_q}{\partial x_j}(\mathbf{y}_q) \right\}, \ j \in \overline{M+2, 3M+3};$$

for $i \in \overline{M+2, 2M+2}$:

$$J_{ij} = \frac{\pi}{\beta} \frac{1}{\tilde{\omega}_{i-M-1}} \sum_{q=1}^{M+1} \left\{ \tilde{\lambda}_{i-M-1,q}^{(-)} \tilde{\omega}_q x_{q+M+1} \frac{\partial \tilde{N}_q}{\partial x_j}(\mathbf{y}_q) \right\}, \ j \in \overline{1, M+1}$$

$\cup \overline{2M+3, 3M+3};$ $J_{ij} = \frac{\pi}{\beta} \frac{1}{\tilde{\omega}_{i-M-1}} \sum_{q=1}^{M+1} \left\{ \tilde{\lambda}_{i-M-1,q}^{(-)} \tilde{\omega}_q x_{q+M+1} \frac{\partial \tilde{N}_q}{\partial x_j}(\mathbf{y}_q) \right\},$

$j \in \overline{M+2, 2M+2};$

for $i \in \overline{2M+3, 3M+3}$:

$$J_{ij} = -\frac{\pi}{\beta} \sum_{q=1}^{M+1} \left\{ \tilde{\lambda}_{i-2M-2,q}^{(+)} - 2\mu^* \theta(\omega_c - |\tilde{\omega}_q|) \right\}$$
$$\times \left( \frac{\partial \tilde{P}_q}{\partial x_j}(\mathbf{y}_q) \cdot \tilde{e}_q(\mathbf{y}_q) + \tilde{P}_q(\mathbf{y}_q) \cdot \frac{\partial \tilde{e}_q}{\partial x_j}(\mathbf{y}_q) \right), \ j \in \overline{1, 2M+2};$$

$$J_{ij} = -\frac{\pi}{\beta} \sum_{q=1}^{M+1} \left\{ \left( \tilde{\lambda}_{i-2M-2,q}^{(+)} - 2\mu^* \theta(\omega_c - |\tilde{\omega}_q|) \right) \right.$$
$$\left. \times \left( \frac{\partial \tilde{P}_q}{\partial x_j}(\mathbf{y}_q) \cdot \tilde{e}_q(\mathbf{y}_q) + \tilde{P}_q(\mathbf{y}_q) \cdot \frac{\partial \tilde{e}_q}{\partial x_j}(\mathbf{y}_q) \right) \right\} - \delta_{ij}, \ j \in \overline{2M+3, 3M+3}.$$

The corresponding expressions for the derivatives of the functions $\tilde{N}_q$, $\tilde{P}_q$ and $\tilde{e}_q$ have the form:

$$\frac{\partial \tilde{N}_q}{\partial x_i}(\mathbf{y}) = -\frac{2}{\pi} \times \begin{cases} \int_{-\mu}^{+\infty} \frac{N_0(E) \cdot x_1}{\left[ (E+x_3)^2 + x_2^2 \tilde{\omega}_q^2 + x_1^2 \right]^2} dE, & i = q \\ \int_{-\mu}^{+\infty} \frac{N_0(E) \cdot x_2 \tilde{\omega}_q^2}{\left[ (E+x_3)^2 + x_2^2 \tilde{\omega}_q^2 + x_1^2 \right]^2} dE, & i = q+M+1 \\ \int_{-\mu}^{+\infty} \frac{N_0(E) \cdot (E+x_3)}{\left[ (E+x_3)^2 + x_2^2 \tilde{\omega}_q^2 + x_1^2 \right]^2} dE, & i = q+2M+2 \\ 0, & \text{otherwise} \end{cases},$$

$$\frac{\partial \tilde{P}_q}{\partial x_i}(\mathbf{y}) = -\frac{2}{\pi} \times \begin{cases} x_1 \int_{-\mu}^{+\infty} \frac{N_0(E) \cdot (E+x_3)}{\left[ (E+x_3)^2 + x_2^2 \tilde{\omega}_q^2 + x_1^2 \right]^2} dE, & i = q \\ x_2 \tilde{\omega}_q^2 \int_{-\mu}^{+\infty} \frac{N_0(E) \cdot (E+x_3)}{\left[ (E+x_3)^2 + x_2^2 \tilde{\omega}_q^2 + x_1^2 \right]^2} dE, & i = q+M+1 \\ \frac{1}{2} \int_{-\mu}^{+\infty} N_0(E) \frac{(E+x_3)^2 - (x_2^2 \tilde{\omega}_q^2 + x_1^2)}{\left[ (E+x_3)^2 + x_2^2 \tilde{\omega}_q^2 + x_1^2 \right]^2} dE, & i = q+2M+2 \\ 0, & \text{otherwise} \end{cases},$$

$$\frac{\partial \tilde{e}_q}{\partial x_i}(\mathbf{y}) = \begin{cases} -\frac{\tilde{\omega}_q x_1 x_2}{\left[ \tilde{\omega}_q^2 x_2^2 + x_1^2 \right]^{3/2}}, & i = q \\ \frac{\tilde{\omega}_q x_1^2}{\left[ \tilde{\omega}_q^2 x_2^2 + x_1^2 \right]^{3/2}}, & i = q+M+1 \\ 0, & \text{otherwise} \end{cases}.$$

The correctness of the above formulas for the Jacobian was also checked numerically (like in Section 3).

**5. Calculated thermodynamic properties of a superconductor.**

In the presented software package, the calculation of the following thermodynamic properties of a superconductor is implemented.

− electron-phonon interaction parameter (dimensionless measure of the strength of $\alpha^2 F(\omega)$) [10]:

$$\lambda_{e-ph} = 2 \int_0^{+\infty} \frac{\alpha^2 F(\omega)}{\omega} d\omega, \tag{14}$$

− superconducting transition temperature in the McMillan approximation, K [10]:

$$T_C = \frac{\langle \omega \rangle}{1.2} \exp\left( -\frac{1.04(1+\lambda_{e-ph})}{\lambda_{e-ph} - \mu^*(1+0.62\lambda_{e-ph})} \right), \tag{15}$$

where $\langle \omega \rangle = \frac{1}{k} \frac{2}{\lambda_{e-ph}} \int_0^{+\infty} \alpha^2 F(\omega) d\omega$,

− Sommerfeld constant, mJ/(mole K$^2$) [13]:

$$\gamma = \frac{2}{3} \pi^2 k^2 N(0)(1+\lambda_{e-ph}), \tag{16}$$

− characteristic phonon frequency, eV [13]:

$$\omega_{\ln} = \exp\left[ \frac{2}{\lambda} \int_0^{+\infty} \ln(\omega) \frac{\alpha^2 F(\omega)}{\omega} d\omega \right], \tag{17}$$

− isotropic superconducting gap at $T = 0$ K (in the McMillan approximation), meV [13]:

$$\frac{2\Delta_0}{k T_C} = 3.53 \left[ 1 + 12.5 \left( \frac{T_C}{\omega_{\ln}} \right)^2 \ln\left( \frac{\omega_{\ln}}{2T_C} \right) \right], \tag{18}$$

– upper critical field at $T = 0$ K, T [13]:

$$\frac{\gamma T_c^2}{H_c^2(0)} = 0.168\left[1 - 12.2\left(\frac{T_c}{\omega_{\ln}}\right)^2 \ln\left(\frac{\omega_{\ln}}{3T_c}\right)\right], \quad (19)$$

– jump in heat capacity at $T = T_c$, J/(mole K) [13]:

$$\frac{\Delta C(T_c)}{\gamma T_c} = 1.43\left[1 + 53\left(\frac{T_c}{\omega_{\ln}}\right)^2 \ln\left(\frac{\omega_{\ln}}{3T_c}\right)\right], \quad (20)$$

– isotopic ratio [13]:

$$\beta = \frac{1}{2}\left(1 - \frac{1.04(1+\lambda)(1+0.62\lambda)}{\left[\lambda - \mu^*(1+0.62\lambda)\right]^2}\mu^{*2}\right). \quad (21)$$

Formula (15) is convenient to use for a preliminary estimate of the temperature range in which $T_c$ is sought.

## 6. Calculations example.

We present the results of calculations performed for the I41/AMD phase of metallic hydrogen under a pressure of 500 GPa for the following parameters values [6]: $\mu^* = 0.1$, $\mu = 6.8$ eV, $\omega_D = 0.34$ eV. Eliashberg function $\alpha^2 F(\omega)$ and the dimensionless electron density of states $N_0(E)$ are shown in figures 2 and 3.

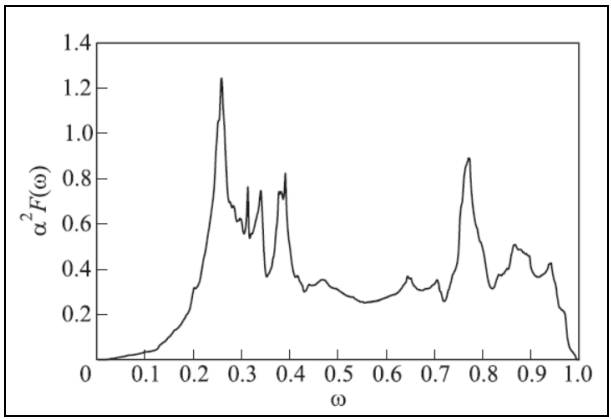

**Figure 2.** The Eliashberg function $\alpha^2 F(\omega)$ for I41/AMD phase of metallic hydrogen under a pressure of 500 GPa [14-16].

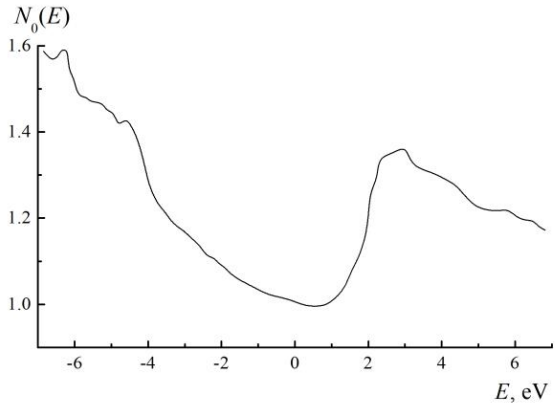

**Figure 3.** The dimensionless electronic density of states $N_0(E)$ for I41/AMD phase of metallic hydrogen under a pressure of 500 GPa [14-16].

The results are the following: $T_c = 221.7$ K (with the correction to the chemical potential of electrons), $T_c = 220.7$ K (without the correction), $\varsigma = 1.68$, $\gamma = 10.2$ mV/(mole K$^2$), $\omega_{\ln} = 0.122$ eV, $\Delta(0) = 42.5$ meV, $H_c(0) = 61.8$ T, $\Delta C(T_c) = 6.36$ J/(mole K), $\beta = 0.487$. The order parameter as a function of temperature is shown in the fig. 4. Although for this compound the difference between the results obtained with and without correction for the chemical potential of electrons is small, for other substances it can be quite significant.

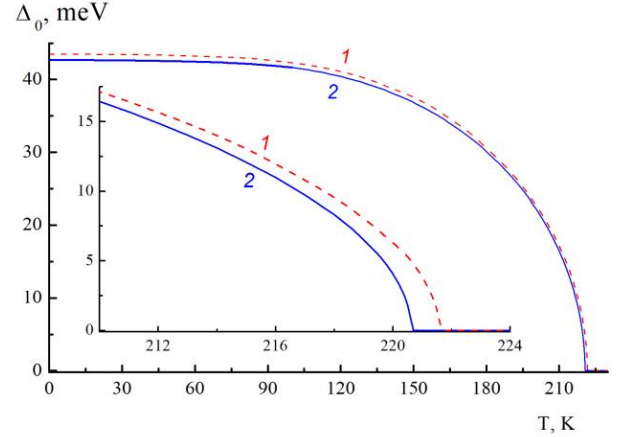

**Figure 4.** The order parameter as a function of temperature $\Delta_0(T)$ for I41/AMD phase of metallic hydrogen under a pressure of 500 GPa (1 – calculations with the correction to the chemical potential of electrons, 2 – without the correction).

Attempts to calculate the temperature of the superconducting transition of metallic hydrogen I41/AMD phase under a pressure of 500 GPa have also been made earlier. So, this temperature calculated in the work [6] by solving the system of Eliashberg equations for the real axis was 217 K. This result is in good agreement with ours, especially given the hypothetical nature of the compound.

In the work [17], the authors presented the calculations of the superconducting transition temperatures for high-temperature superconductors with hydrogen-based alloy backbone at various pressures. For all substances, for which the Eliashberg functions in [17] are given, we calculated $T_c$ by two methods (see Table 1): using the Allen-Dynes algorithm (column "A-D") and solution of the Eliashberg equations of the form (6) (column "El-1"). $T_c$ obtained in [17] by solving the Eliashberg equations are also given in the table (column "El-2").

**Таблица 1.** Superconducting transition temperatures of various substances with hydrogen-based alloy backbone at corresponding pressures (GPa).

| Substance | P, GPa | $T_C$, K | | |
|---|---|---|---|---|
| | | **A-D** | **El-1** | **El-2** |
| LaBeH$_8$ | 50 | 194 | 192 | 191 |
| LaBH$_8$ | 70 | 160 | 159 | 160 |
| LaAlH$_8$ | 100 | 145 | 145 | 144 |
| CaBeH$_8$ | 210 | 307 | 303 | 302 |
| CaBH$_8$ | 100 | 241 | 239 | 238 |
| YBeH$_8$ | 100 | 250 | 248 | 249 |
| SrBH$_8$ | 150 | 206 | 204 | 200 |

As can be seen from Table 1, firstly, the results of the calculations of our program and the program used by the authors of [17] mainly coincide with an accuracy of 1 K (the discrepancy is obviously due to the error in digitizing the graphs of the Eliashberg functions). Secondly, calculations by the Allen-Dynes algorithm, in this case, give a very good approximation to the superconducting transition temperature (no worse than 2 K for most materials).

## 7. Conclusions.

In our work, we present a software package for solving the Eliashberg system of equations written in two different forms on the imaginary axis. Also, we present the results of calculations of the superconducting transition temperature and thermodynamic properties of the metallic hydrogen I41/AMD phase under a pressure of 500 GPa. In addition, $T_c$ for various substances with hydrogen-based alloy backbone at various pressures are presented. In all cases, the superconducting transition temperatures are in good agreement with the results of the original papers.

As a natural generalization of the results presented in this work, the model based on Eliashberg equations could be expanded to take into account the influence of impurities and the multiband nature of some compounds. It could also be expanded to take into account the possibility of the inhomogeneous states formation (see, for example, [18, 19]) that arises when approaching a first-order phase transition between liquid and crystalline metallic hydrogen.

In addition to the Eliashberg approach, we also plan to take into account the possibility of the formation of two Bose condensates in the system, the first of which is a Bose condensate of bielectron Cooper pairs, and the second is a biproton Bose condensate in the ionic component [20, 21].

**Declaration of competing interest**

The authors declare that they have no known competing financial interests or personal relationships that could have appeared to influence the work reported in this paper.

**Acknowledgment**

M. Yu. K. and R. Sh. I. gratefully acknowledge a support of the RFBR (grant № 20-02-0015) and a support from the Basic Research Program of the National Research University Higher School of Economics.